\documentstyle[epsfig,12pt]{article}    
\newcommand{\eeq}{\end{equation}}
\newcommand{\beq}{\begin{equation}}

\begin{document}

\title{On the Dalitz Plot Approach in Non-leptonic Charm Meson Decays}

\author{I. Bediaga, C. G\"obel \\
Centro Brasileiro de Pesquisas F\'\i sicas, R. Dr. Xavier Sigaud 150\\
22290 -- 180 --  Rio de Janeiro, RJ, Brazil\\
and\\
R. M\'endez--Galain\\
 Instituto de F\'{\i}sica, Facultad de Ingenier\'{\i}a,\\
CC 30, CP 11000 Montevideo, Uruguay}

\date{ \ \today }

\maketitle

\begin{abstract}
We claim that the non-resonant contribution to non-leptonic charm 
meson decays may not be constant in the phase space of
the reaction. We argue that this can be 
relevant for any weak reaction. 
We discuss in detail the decay $D^+ \to K^- \pi^+ \pi^+$.

\end{abstract}

Non-leptonic charm meson decays have been extensively studied 
both theoretically and experimentally. The high diversity and low
multiplicity of decay channels provide important information on both
weak and strong interactions. These decays have contributions from resonances 
in intermediate states, as well from the direct non-resonant (NR) decay. 
The understanding of the decay pattern of charm mesons
as a whole, and therefore the extraction of the decay partial widths for all 
contributing states, is essential in addressing many open problems in charm 
physics.

The Dalitz plot analysis \cite{dalitz} is a powerful technique widely
used in the study of resonance substructures
on charmed meson decays. The plot represents the phase space 
of the decay and it is weighted by the squared amplitude of the reaction. 
Therefore, it contains information on both the kinematics and the dynamics. 
Within this technique, intermediate resonant and non-resonant
contributions are fitted to get the respective amplitudes and phases.
The corresponding partial decay widths can then be obtained.

 When experimental data on 
non-leptonic decays of charm mesons became available in the seventies, 
J. Wiss {\it et al} \cite{wiss} used the Dalitz plot technique to search for the
spin of the recently discovered charged $D$ meson. They  found a result 
statistically compatible with a flat distribution.  Assuming that the 
structure on the Dalitz plot is dominated by the hadronic spin amplitude
\cite{book}, they concluded  the $D^+$ meson would be a spin 0 particle. 

Subsequently, resonances were found in higher statistics experiments. Since
then, attention has focused on them and the NR contribution has been assumed
to be constant. For instance, data on non-leptonic decays of the $D$ meson has
been fitted \cite{mark2,mark3,691,argus,687} using Breit-Wigner 
functions\cite{jackson}  to represent the various resonances (with the
respective angular distribution) and a constant function to describe the NR 
contribution \cite{foot0}.

Although the above parameterization is widely used, a very poor fit has been
reported \cite{mark3,687}, suggesting that it may not be adequate to describe
these decays. These poor results do not improve with higher statistics
or considering a larger number of resonances\cite{687}. 
Moreover, this problem appears in all the $D \to {K} \pi \pi$
decay channels already measured
($D^0 \to \bar{K}^0 \pi^+ \pi^-$, $D^+ \to \bar{K}^0 \pi^+ \pi^0$,
$D^+ \to K^- \pi^+ \pi^+$ and $D^0 \to K^- \pi^+ \pi^0$) \cite{foot} and 
the worst
fit is obtained for  $D^+ \to K^- \pi^+ \pi^+$, where
the NR contribution dominates\cite{687}. (In this case, with
29 degrees of freedom, the $\chi^2$ per degree of freedom is as bad as 3.01.)

A possible explanation for these discrepancies is the incorrect use of 
a constant amplitude for the NR contribution. An incorrect parameterization will certainly influence the fit
of the resonances and consequently the extracted values of amplitudes
and phases. As an example, MarkIII reported\cite{mark3}
significant discrepancies on the measurement of the branching ratio (BR) of
$D^+ \to {\bar K}^* \pi^+$ obtained from the different final states 
$K^0 \pi^0 \pi^+$ and $K^- \pi^+ \pi^+$.
Note that while the NR contribution to the first final state is
of the order of 15\% of the total partial decay width, in the second it is as
large as 80\%.

Here, we claim that NR charm meson decays may contain information beyond
the simple hadronic amplitude of a spin zero particle decaying into three
spin zero daughters. Since we are dealing with weak decays, 
signatures of this fundamental interaction can directly
appear in the NR amplitude. In weak interactions between quarks 
and leptons helicity plays an important role. Consequently, one expects a 
significant dependence of the weak amplitudes on the momenta of the 
interacting particles. Thus, the dynamics of these reactions vary
 from  point to point of the phase space and the significance of this variation
depends on the specific physical reaction.

This should be particularly important in weak decays of charm mesons. 
The large value of the charm quark mass allows for a
quasi perturbative treatment of QCD. Furthermore, charm quark decays
into light quarks and this enhances the importance of helicity. 
For example, we can see the effect of weak partonic mechanism
responsible for the Cabibbo favored  $D$ meson decays, i.e. $c \to s u \bar{d}$,
by analysing the decay of $\tau$ leptons, $\tau \to \mu \bar{\nu}_{\mu} 
\nu_{\tau}$, which are essentially similar. This simple example will shed some 
light on the dependence of a weak reaction on its phase space.

The theoretical Dalitz plot corresponding to the decay  $\tau \to \mu 
\bar{\nu}_{\mu} \nu_{\tau}$ can be obtained by taking the well known
decay amplitude of pure leptonic decays\cite{okun}. This decay amplitude
 can be written as a function of two invariant variables defining
a Dalitz plot, e.g.,  $m_{\mu 
\bar{\nu}_{\mu}}^2 \equiv (p_{\mu} 
+ p_{\bar{\nu}_{\mu}})^2$
and $m_{\mu \nu_{\tau}}^2 \equiv (p_{\mu} + p_{\nu_{\tau}})^2$ to give
\begin{equation}
|{\cal M }_{\tau \to \mu 
\bar{\nu}_{\mu} \nu_{\tau}}|^2 \propto m_{\mu \nu_{\tau}}^2 (m^2_{\tau} 
- m_{\mu \nu_{\tau}}^2)
\label{tau}
\end{equation}
where $m_{\tau}$ is the $\tau$ mass.

The dynamics of the reaction has a quadratic dependence on
the variable  $m_{\mu \nu_{\tau}}^2$. 
As the
Dalitz plot is weighted by $|{\cal M }_{\tau \to \mu 
\bar{\nu}_{\mu} \nu_{\tau}}|^2 $, equation (\ref{tau}) shows
that a Dalitz plot of a pure
weak decay has indeed significant variations along the phase space.

Obviously, due to the hadronization process of the partons after their weak 
interaction, the result of the previous example cannot be simply translated into
hadronic decays. In the latter case, one has to take into account 
non-perturbative QCD effects involved in the final hadronic state formation.
In order to make an estimate of the effect of the dynamics in the Dalitz plot, 
we use an approximate method to describe hadronic decays.
The method is based on both the factorization technique \cite{fal}
and an effective Hamiltonian \cite{bsw,buras}
for the partonic interaction and has been successfully used to describe 
heavy meson decays\cite{fal,more}.

\begin{figure}
\begin{center}
\mbox{\epsfig{file=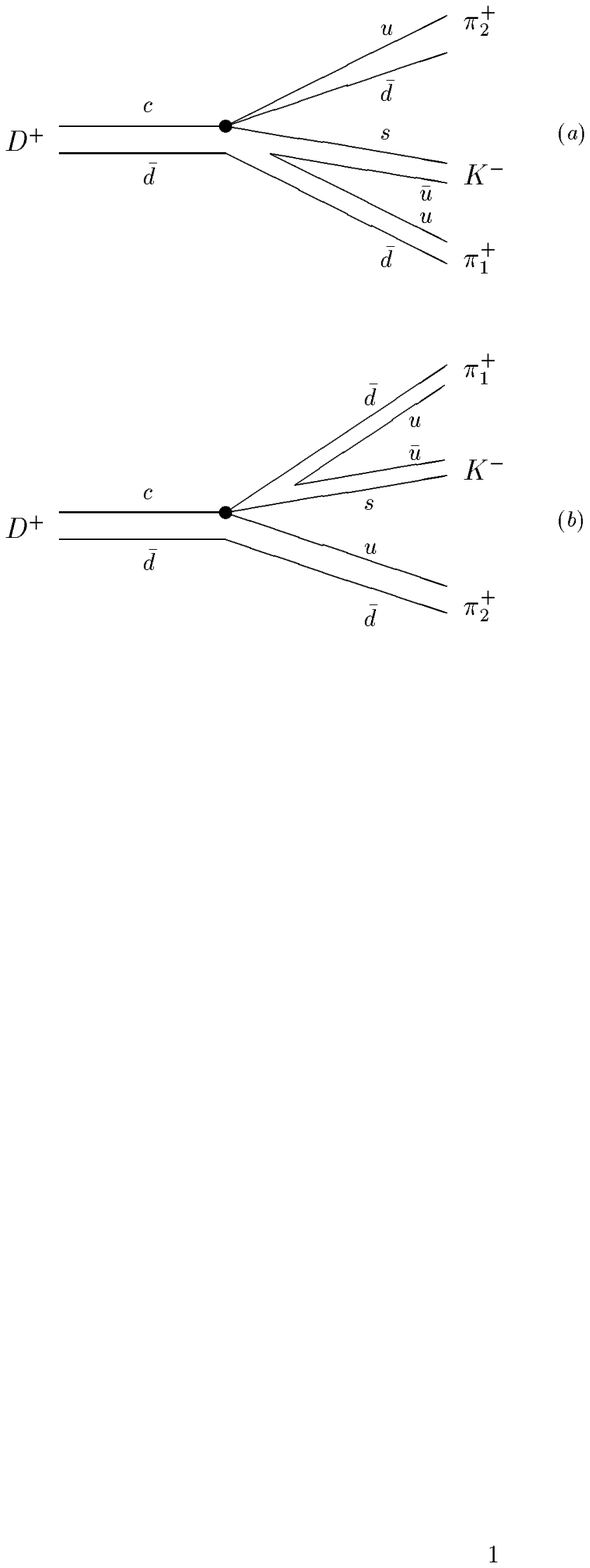,width=8cm}}
\caption{The two diagrams contributing to the decay $D^+ \to K^- \pi^+ \pi^+$
according to the effective Hamiltonian of equation (2).}
\end{center}
\label{diagrm}
\end{figure}

As we are interested in the NR contributions,
we analyse the channel $D^+ \to K^- \pi^+ \pi^+$, which has a
very large NR branching ratio, as mentioned above. The effective Hamiltonian
for the weak vertex $c \to s u \bar{d}$ is \cite{bsw,buras}:
\begin{equation}
{\cal H}_{eff} = (\frac{G_F}{\sqrt{2}}) \cos ^2 \theta_c [ a_1 
:(\bar{s}c)(\bar{u}d): + a_2 :(\bar{s}d)(\bar{u}c):]
\label{heff}
\end{equation}
where $(\bar{q}q')$ is a short-hand notation for $\bar{q} \gamma^{\mu} 
(1-\gamma_5)q'$. The 
coefficients
$a_1$ and $a_2$ characterize the contribution of the effective charged and neutral 
currents respectively,
which include short-distance QCD effects. Their values have been fitted in the case 
of charm meson two-body decays (see for example reference \cite{bsw}). The diagrams 
contributing to the decay $D^+ \to K^- \pi^+ \pi^+$ are shown in Figure 
(\ref{diagrm}). Using factorization we obtain the following decomposition for the 
hadronic amplitude 
\begin{eqnarray}
{\cal M }_{D^+ \to K^- \pi^+ \pi^+} & = &
 (\frac{G_F}{\sqrt{2}}) \cos ^2 \theta_c [ a_1 \langle K^-\pi_1^+|\bar{s}c|D^+
\rangle \langle \pi_2^+|\bar{u}d|0\rangle 
\nonumber \\
& & + a_2 
\langle K^-\pi_1^+|\bar{s}d|0\rangle\langle \pi_2^+|\bar{u}c|D^+\rangle 
\: \: + \: \:  (\pi_1^+ \leftrightarrow  \pi_2^+) ] ~.
\label{mdkpp}
\end{eqnarray}

Let us first discuss the term driven by $a_1$, i.e, the one of Figure 
(\ref{diagrm}.a). The most general form to decompose the first matrix 
element can be written in terms of four form factors\cite{marshak}.
Using the parameterization of reference \cite{kuhn}, we can write:
\begin{equation}
\langle K^-\pi_1^+|\bar{s}c|D^+\rangle = A_1^{\mu} F_1 + A_2^{\mu} F_2 + 
i V_3^{\mu}  F_3 + A_4^{\mu} F_4~,
\label{a11}
\end{equation}
where
\begin{eqnarray}
 & & \mbox{}A_1^{\mu} = p_K^{\mu} + p_D^{\mu} - Q^{\mu} \frac { Q \cdot 
( p_K + p_D) } {Q^2}~, \nonumber \\
 & & \mbox{}A_2^{\mu} = p_{\pi_1}^{\mu} + p_D^{\mu} - Q^{\mu} \frac { Q 
\cdot ( p_{\pi_1} + p_D) } {Q^2}~, \nonumber \\
 & & \mbox{}V_3^{\mu} = \epsilon^{\mu\alpha\beta\gamma} p_K^{\alpha} 
p_{\pi_1}^{\beta} p_D^{\gamma}~,
\nonumber \\
 & & \mbox{}A_4^{\mu} = Q^{\mu} = p_K^{\mu} + p_{\pi_1}^{\mu} - p_D^{\mu} 
= -p_{\pi_2}^{\mu} ~.
\nonumber
\end{eqnarray}
The terms proportional to $F_1$, $F_2$ and $F_4$ originate from the axial 
vector part of the matrix element
whereas the one proportional to $F_3$ originates from the vector part; 
the terms proportional to $F_1$, $F_2$ and $F_3$ correspond to spin 1 and $F_4$
to spin 0. The four form factors depend on three variables $m_1^2 = 
(p_k + p_{\pi_1})^2$, $m_2^2 = (p_k + p_{\pi_2})^2$ and $Q^2$ which is a constant
($m^2_{\pi}$) in this case.

The second matrix element in equation (\ref{mdkpp}) has the well known form
\begin{equation}
\langle \pi_2^+|\bar{u}d|0\rangle = i f_{\pi} p^{\mu}_{\pi_2} ~.
\label{pion}
\end{equation}

The only contributing term in equation (\ref{a11}) after multiplying it 
by equation (\ref{pion}), is the axial spin 0 term, i.e.,
\begin{equation}
\langle K^-\pi_1^+|\bar{s}c|D^+\rangle \langle \pi_2^+|\bar{u}d|0\rangle = 
   (p_{\pi_2~\mu} \; F_{4}) \: ( i f_{\pi} {p_{\pi_2}}^{\mu})  = 
 i f_{\pi}   m^2_{\pi} F_{4} ~.
\label{a1}
\end{equation}
 
To find the contribution of Figure (\ref{diagrm}.b), one 
can use the well known expressions\cite{bsw2}
\begin{eqnarray}
\langle \pi_2^+|\bar{u}c|D^+\rangle & = & 
\left[ (p_D + p_{\pi_2})^{\mu} - \frac{m^2_D - m^2_{\pi}}{q^2}
(p_D - p_{\pi_2})^{\mu} \right]  F^{1^-}_{D\pi}(q^2) \nonumber \\
&  & + \: \frac{m^2_D - m^2_{\pi}}{q^2}
(p_D - p_{\pi_2})^{\mu} F^{0^+}_{D\pi}(q^2)
\nonumber
\end{eqnarray}
and
\begin{eqnarray}
 \langle K^- (p_K) \pi_1^+|\bar{s}d|0\rangle & = &
\langle \pi_1^+|\bar{s}d|K^+ (-p_K) \rangle = \nonumber \\
& & \left[ (-p_K + p_{\pi_1})^{\mu} - \frac{m^2_K - m^2_{\pi}}{q^2}
(-p_K - p_{\pi_1})^{\mu} \right]  f_+(q^2) \nonumber \\
& & + \frac{m^2_K - m^2_{\pi}}{q^2} 
(-p_K - p_{\pi_1})^{\mu} f_0(q^2) ~.
\nonumber
\end{eqnarray}

In the equations above, $q^2 = (p_D - p_{\pi_2})^2 = (-p_K-p_{\pi_1})^2$
 while the functions $F^{J^P}_{D\pi}(q^2)$ (corresponding to a current of spin
parity $J^P$), $f_+(q^2)$ and $f_0(q^2)$ are form factors. We return to them later.

We then find for the second contribution in equation (\ref{mdkpp}),
\begin{eqnarray}
& & \langle \pi^+|\bar{u}c|D^+\rangle \langle K^- \pi^+|\bar{s}d|0\rangle 
= F^{1^-}_{D\pi}(m_1^2) f_+(m_1^2) \: (m_D^2 + m_K^2 + 
2 m_{\pi}^2 - 2 m_2^2
-m_1^2) \nonumber \\
& & \:\:\: + \:\:[ F^{1^-}_{D\pi}(m_1^2) f_+(m_1^2) - F^{0^+}_{D\pi}(m_1^2) 
f_0(m_1^2) ]\:  \frac{(m_D^2 - m_{\pi}^2)(m_K^2 - m_{\pi}^2)}{m_1^2}
\nonumber \\
& &  \:\:\: + \:\: (m_1^2 \leftrightarrow m_2^2)
\label{a2}
\end{eqnarray}
where we have explicitly introduced the Dalitz plot variables $m_1^2$
 and  $m_2^2$ defined above.
 
The contribution of diagram (\ref{diagrm}.a), given by equation (\ref{a1}) 
is proportional to $f_{\pi} m_{\pi}^2 $. Thus, unless the form factor $F_4$
is   unacceptably large ($F_4 \sim 10^3$), we can safely neglect 
this contribution in favor of that of diagram (\ref{diagrm}.b), given by equation 
(\ref{a2}) which contains $m_D^2$. As an aside, it is possible that
the NR part of the decay $D^+ \to K^- \pi^+ \pi^+$ is large
precisely because the contribution of  diagram (\ref{diagrm}.b) is {\it not} small.

The NR contribution to the amplitude of the decay
$D^+ \to K^- \pi^+ \pi^+$ can thus be simply written replacing equation
(\ref{a2}) in (\ref{mdkpp}), neglecting the contribution of Figure (\ref{diagrm}.a). 
The final expression thus depends on the effective coefficient 
$a_2$ and the four form factors.
The two $D\pi$ form factors $F^{J^P}_{D\pi}(q^2)$, have well established 
expressions\cite{buras}~:
\begin{equation}
F^{J^P}_{D\pi}(q^2) = \left(1-\frac{q^2}{M^2_{D\pi,J^P}}\right)^{-1} 
\label{ff}
\end{equation}
where $M_{D\pi,1^-} = 2.01$ GeV and  $M_{D\pi,0^+} = 2.2$ GeV. 
They have been successfully used in the kinematic range we are
considering here. The poles lie outside our kinematic region.
The $K\pi$ form factors, $f_+(q^2)$ and $f_0(q^2)$, can be extracted from
the semi-leptonic decays $K\to \pi l \nu$, with $l=e, \mu$. 
Nevertheless, it is not clear that
the usual parameterization\cite{okun}
\begin{equation}
f_+(q^2) = f_+(0) \left(1+ \lambda_+ \frac{q^2}{m^2_{\pi}}\right)~, ~
f_0(q^2) = f_0(0) \left(1+ \lambda_0 \frac{q^2}{m^2_{\pi}}\right)
\label{fkpi}
\end{equation}
is valid in the  whole kinematic region of our reaction. 
In equation (\ref{fkpi}), 
$f_+(0)=f_0(0)=1$ and the other coefficients have been measured to be\cite{pdg}:
$\lambda_+\approx 0.03$ independent of the measured channel,  
whereas the value of $\lambda_0$ depends on the decay:  $\lambda_0 \approx 0$ for 
$K^-\to \pi^0 \mu^- \nu$ and $\lambda_0 \approx 0.025$ for $K^0\to \pi^+ \mu^- \nu$.

In order to check the validity of this calculation scheme, we have evaluated the
NR partial 
decay width $\Gamma (D^+ \to K^- \pi^+ \pi^+)_{NR}$ ~ using the expressions above.
With $\lambda_0=0$ and the value of
$a_2$ extracted from two body decay\cite{bsw}, we find a branching ratio (BR) of
9\% which is close to the reported 
experimental value\cite{pdg} $7.3\pm 1.4$\% obtained by fitting the NR contribution to 
a constant. We studied the stability 
of this result under the change of the parameters $\lambda_+$ and $\lambda_0$~:
if we take the various values
extracted from different channels we find that 
the BR varies less than 30\%. Even assuming constant form factors ($\lambda_+=\lambda_0=0$), 
the BR remains of the same order of magnitude.

\begin{figure}
\begin{center}
\mbox{\epsfig{file=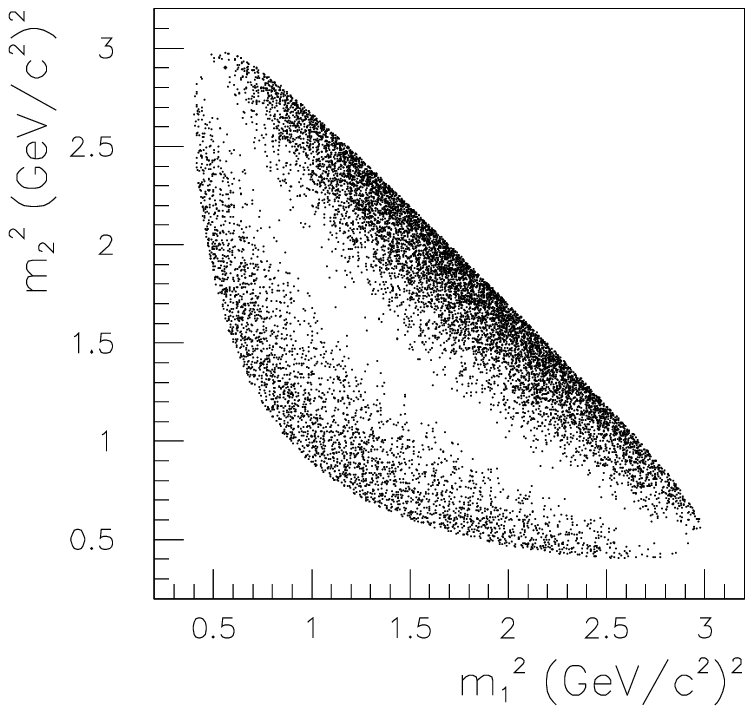,width=15cm}}
\caption{The Dalitz plot of the decay $D^+ \to K^- \pi^+ \pi^+$, weighted by
$|{\cal M}_{D^+ \to K^- \pi^+ \pi^+}|^2$ as in equations (3) and (7),
generated via Monte Carlo. The Dalitz plot variables are 
$m_1^2 \equiv (p_K +p_{\pi_1})^2$ and $m_2^2 \equiv (p_K +p_{\pi_2})^2$.}
\end{center}
\label{Dfig}
\end{figure}

Figure (\ref{Dfig}) shows the Dalitz plot for the NR contribution to 
the decay
$D^+ \to K^- \pi^+ \pi^+$ as a function of the variables $m_1^2$ and $m_2^2$.
It has been generated by Monte Carlo simulation with a weight 
 proportional to the square of the amplitude in equation
(\ref{mdkpp}), using equation (\ref{a2}). We have considered the
same central value of the parameters as above.
As one can see from equation (\ref{a2}) and Figure (\ref{Dfig}), 
according to this calculation the  matrix element describing the dynamics
of the NR contribution to the decay $D^+ \to K^- \pi^+ \pi^+$
significantly varies along the phase space of the reaction.
Its shape remains almost the same for other values of the
parameters of the $K\pi$ form factor. This is still valid even if we take 
the four form factors as constants.

However,  the result presented in Figure (\ref{Dfig}) has been obtained 
using an approximate method. Non-perturbative effects, present 
in this decay through the exchange of soft gluons, or final state 
interactions could change the structure shown 
in this figure. In the extreme case where non-perturbative 
effects completely dominate the decay, the structure will be
washed out because of the dispersive nature of these effects,
therefore obtaining the flat contribution predicted by the pure hadronic 
decay of a zero spin particle. Comparison between the distribution shown 
in Figure (\ref{Dfig}) and experimental data will thus be a test 
for the validity of the factorization method.

In summary, we have shown that the natural parameterization for the 
non-resonant part of charm decays -- based in the spin amplitude of 
the hadronic decay -- could significantly change due to the fundamental 
weak interaction between quarks. The appearance of these structures in the plot
could be responsible for the problems of the various experimental
teams with the convergence of their fits. To clarify this point, it is 
important in future analyses to use a parameterization for the non-resonant 
contribution going beyond the simple constant. 

We acknowledge Angela Olinto and Alberto Reis for useful discussions. One of us (RMG) wants
to thank the warm hospitality at LAFEX-CBPF. Two of us (IB, CG) want to
thank the CNPq (Brazil) for financial support.

\newpage

\begin{thebibliography}{999}
\bibitem{foot0} In fact, Mark III tried to fit using a different
parameterization for the NR amplitude, since they were having problems 
with their fit.
\bibitem{foot} Throughout this text, charge conjugate states are implied.
\end{thebibliography}
\end{document}